# Plasmonic finite-thickness metal-semiconductor-metal waveguide as ultra-compact modulator

**Viktoriia E. Babicheva**[*], **Radu Malureanu**, and **Andrei V. Lavrinenko**

*Department of Photonics Engineering, Technical University of Denmark, Ørsteds Plads, Bld. 343, DK-2800 Kongens Lyngby, Denmark*

* Corresponding author (V.E. Babicheva). Present address: Purdue University, Birck Nanotechnology Center, 1205 West State Street, West Lafayette, Indiana, 47907-2057 USA. E-mail: vbab@fotonik.dtu.dk, baviev@gmail.com

**Abstract:** We propose a plasmonic waveguide with semiconductor gain material for optoelectronic integrated circuits. We analyze properties of a finite-thickness metal–semiconductor–metal (F-MSM) waveguide to be utilized as an ultra-compact and fast plasmonic modulator. The InP-based semiconductor core allows electrical control of signal propagation. By pumping the core we can vary the gain level and thus the transmittance of the whole system. The study of the device was made using both analytical approaches for planar two-dimensional case as well as numerical simulations for finite-width waveguides. We analyze the eigenmodes of the F-MSM waveguide, propagation constant, confinement factor, Purcell factor, absorption coefficient, and extinction ratio of the structure. We show that using thin metal layers instead of thick ones we can obtain higher extinction ratio of the device.

## 1. Introduction

Surface plasmon polaritons (SPPs) are surface waves at the interface between metal and dielectric media. Due to their tight binding at the interface, they can be used for controlling light at nanoscale levels [1]. Plasmonic devices are promising candidates for photonic integrated circuits because of their small footprints and high operation speed [2-5].

Various types of plasmonic waveguides have been analyzed. Stripe waveguides based on a single thin metal film provide a relatively high propagation length, i.e. low modal losses, but suffer from low field confinement [6]. As a consequence, there was a turn toward hybrid plasmonic waveguides that combine thin metal stripes with semiconductor and dielectric layers of different refractive index [7-14]. Furthermore, these waveguides were improved by adding gain material and thus making possible to mitigate the modal losses [15-19].

From another side, a structure where a dielectric core is sandwiched between two metal layers is very promising as it can provide even smaller dimensions and higher field localization [20,21]. A metal–insulator–metal (MIM) waveguide does not exhibit cutoff even at very small core thickness and as the result allows unprecedented thin layouts of tens of nanometers [22-24]. Planar MIM plasmonic waveguides with thin metal claddings were studied and optimized for various purposes [25-29]. They possess several symmetric and asymmetric modes as well as metal-clad and quasi-bound ones, depending on the frequency range [25].

Based on these waveguides different types of plasmonic modulators have been studied during last few years [30,31]. It has been shown that some designs outperform conventional silicon-based modulators [32,33]. A plasmonic waveguide based on an active core sandwiched between thick metal plates has promising characteristics [31,34-36]. Apart from the speed and dimensions advantages, the metal plates can serve as electrodes for electrical pumping of the active material making it easier to integrate. Including an additional thin layer in a MIM waveguide, e.g. silicon oxide, vanadium oxide or transparent conductive oxide, allows the control of the dispersive properties of the waveguide and thus increasing the efficiency of a plasmonic modulator [24,37-42].

Recently, it was proposed to utilize a metal–semiconductor–metal (MSM) structure with the InP-based semiconductor gain core as a plasmonic modulator [35]. Bulk semiconductor medium, quantum wells and quantum dots as active material were studied for different layouts of the modulator. The analysis shows that an extinction ratio of several decibels per micron can be achieved, depending on the gain level.

In this paper we combine both characteristic features and propose a finite-thickness MSM (F-MSM) structure with the active semiconductor core sandwiched between thin metal layers. In Section 2 we compare characteristics of all three eigenmodes in a F-MSM waveguide specifying the best one to be exploited in the plasmonic modulator. Analytical calculations of a F-MSM relative effective index and field confinement factors are presented in Section 3. In Section 4 we show subsequent increasing of the absorption coefficient and possibility to control wave propagation. In Section 5 influence of n- and p-doped layers is analyzed. Sections 6 and 7 are dedicated to numerical simulations for finite-length and finite-width waveguides, respectively. We analyze and summarize results in Section 8.

## 2. F-MSM waveguide modes

As the first step we determine the eigenmodes in a F-MSM waveguide. We solve the SPP dispersion equation for the five-layer system that corresponds to the two-dimensional (2D) F-MSM waveguide shown schematically in Fig. 1a. Calculations are performed for the telecommunication wavelength 1.55 μm. The InP-based gain core has different thickness $d$ and permittivity $\varepsilon = 12.46 + i\varepsilon'' = (n' + in'')^2$, where $n'$ and $n''$ are the real and imaginary parts of the refractive index, respectively. The imaginary part of the permittivity can be defined as $\varepsilon'' = -gn'/k_0$, where $g$ is the material gain and $k_0$ is the free-space wave-number. We analyzed the system with various metal-layer thicknesses $t$ assuming a metal permittivity $\varepsilon_{Ag} = -128.7 + 3.44i$, corresponding to the one of silver at this frequency. The silica layers with permittivity $\varepsilon_d = 2.34$ are assumed to extend infinitely in the z-direction (see Fig.1a). Further, we introduce the relative effective index $n_{eff} = \beta/(n'k_0)$, where $\beta$ is the SPP propagation constant. The relative effective index is defined as the ratio between the effective mode index and the material refractive index. In our case, material InP has refractive index $n'_{InP} = 3.53$.

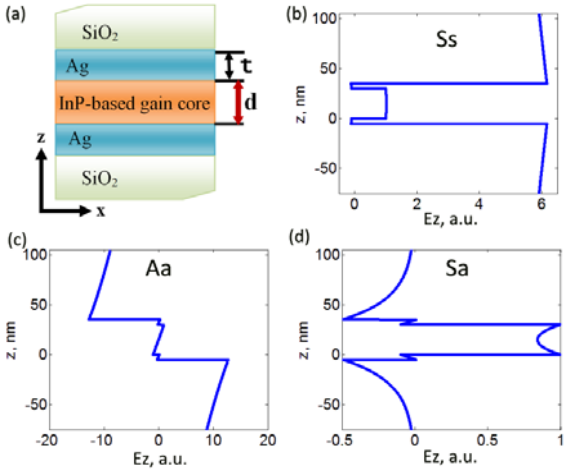

Fig. 1. (a) Finite-thickness MSM waveguide. (b), (c), (d) $E_z$ field profile in the waveguide with $d = 30$ nm and $t = 5$ nm for Ss, Aa and Sa modes, respectively. The field is normalized such that $E_z = 1$ a.u. at the interface between gain core and metal layer.

A F-MSM waveguide with $d = 30$ nm and $t = 5$ nm supports three modes (Fig. 1). Two modes have symmetric electric field inside the core and symmetric/asymmetric fields inside the metal layers (notation Ss and Sa, respectively). The third mode has asymmetric fields inside all layers (Aa mode). The mode with the asymmetric field inside the core and symmetric one in the metal layers (As mode) is not supported by the waveguide with such thin core.

The characteristics of the modes are summarized in Table 1. It can be seen that the absorption coefficient α increases together with increasing of the propagation constant β. The field of the Ss mode is mostly located outside the waveguide core, residing mostly in the low loss dielectric surrounding the waveguide. Due to this, the absorption coefficient of the mode is much lower than the ones for Aa and Sa modes. Furthermore, the modes that possess the highest absorption also have the most pronounced response Δα when changing the gain of the semiconductor core. Characteristic value $q$ shows at which distance the field amplitude drops $e$ times and can be found as $q = 1/\text{Im}(k_z)$, where $k_z = (\varepsilon_d k_0^2 - (\beta+i\alpha)^2)^{1/2}$ is the z-component of the propagation vector.

The field confinement factor Γ can be defined with respect to either field intensity $E^2$ or Poynting vector component $P_x$ along the propagation direction. Also, based on how the waveguide is defined, it can be calculated as the ratio of the electric field energy (flux) inside the core (notation "core") or electric field energy (flux) confined inside the core and both metal plates (notation "core + 2Me") to the total amount of electric field energy (flux) The confinement of Ss and Aa modes is very low due to the field extension outside the waveguide. It is also fully consistent with rather weak response Δα and large characteristic distance $q$.

The highest confinement of the Sa mode leads to the largest response Δα. So, our further analysis of the F-MSM waveguide as plasmonic modulator is performed only for the Sa mode, the one showing the best ability to be controlled [43].

Table 1. Characteristics of different modes in the F-MSM waveguide with $d = 30$ nm and $t = 5$ nm

| Type | Ss | Aa | Sa |
|---|---|---|---|
| $\beta$, μm$^{-1}$ | 6.22 | 8.11 | 42.88 |
| $\alpha$, μm$^{-1}$ | 1.8x10$^{-4}$ | 0.127 | 0.59 |
| $\Delta\alpha = \alpha_{off} - \alpha_{on}$, μm$^{-1}$ ($g_{on} = 2000$ cm$^{-1}$) | 1.16 x10$^{-4}$ | 0.0198 | 0.32 |
| $q$, μm | 1.86 | 0.19 | 0.024 |
| $\Gamma$ ($E^2$, core) | 5.0 x10$^{-4}$ | 0.044 | 0.65 |
| $\Gamma$ ($E^2$, core + 2Me) | 5.4 x10$^{-4}$ | 0.058 | 0.72 |
| $\Gamma$ ($P_x$, core) | 2.5 x10$^{-3}$ | 0.0017 | 0.95 |
| $\Gamma$ ($P_x$, core + 2Me) | 2.6 x10$^{-3}$ | 0.0020 | 0.96 |

## 3. Field confinement in the F-MSM waveguide

One possibility to increase the effective index is to reduce the metal thickness $t$ [44]. We studied properties of waveguides taking several thicknesses of the core ($d = 30…200$ nm) and metal films. In the latter case we investigated two routes. One is changing the thickness of both metal films simultaneously. Another one is to fix the thickness of one film ($t_1 = 70$ nm) and vary only the thickness of the second film.

The structures with relatively thick metal layers possess modal properties similar to a 3-layered infinite MSM waveguide: the relative effective index increases with decreasing of *d* and almost does not depend on the metal thickness (Fig. 2a). However, for thin metal layers the relative effective index shows a strong dependence on the metal thickness. Relative effective index $n_{eff}$ increases several times for small metal thicknesses $t < 15$ nm.

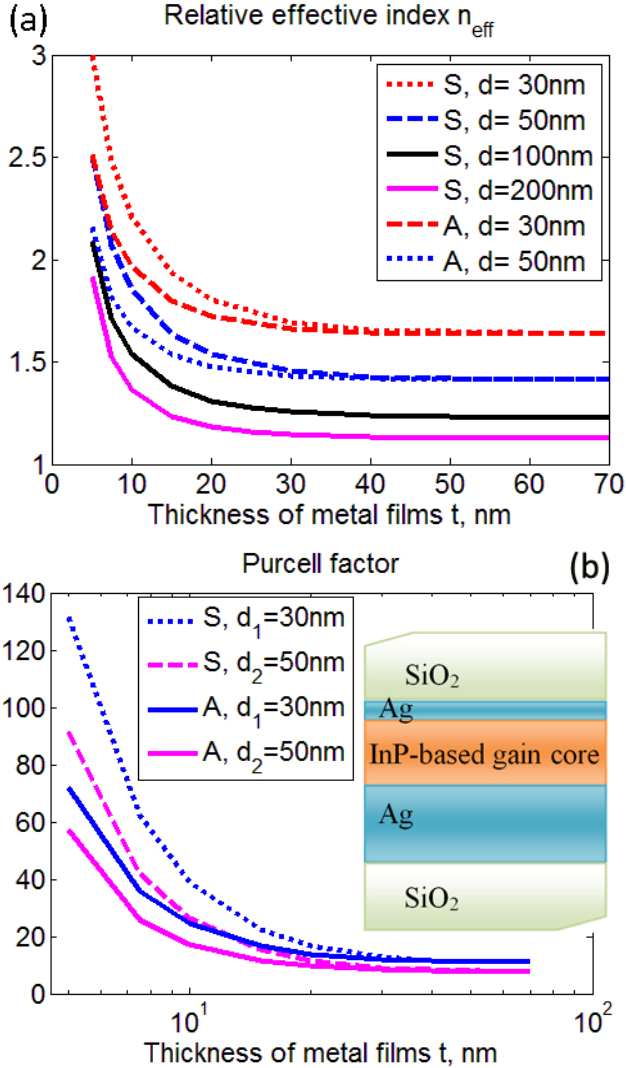

Fig. 2. Relative effective indices (a) and Purcell factor (b) of F-MSM waveguides significantly increase for thin metal layers. "S" corresponds to the case when thickness of both metal films is varied simultaneously (symmetric structure); "A" corresponds to the case when thickness of one metal film is fixed ($t_1 = 70$ nm) and another varied (asymmetric structure, shown on the inset of (b)).

The trend can be explained in the following way. The F-MSM waveguide with thin metal layers can be considered as two coupled metal stripe waveguides. Each metal strip supports an asymmetric mode often referred to as the short-range SPP (see Fig. 1d). This mode has higher effective index and absorption coefficient than the long-range SPP and its effective index dramatically increases with the decrease of the metal thickness [6,45]. Therefore, the effective index of two coupled stripe waveguides is substantially increased (see Fig. 2a).

In the case of an active MSM waveguide, we can define the Purcell factor (PF) to show the increase of the radiative rate. It was shown that MIM and MSM waveguides with active materials possess quite high PF [46-49]. Moreover, the PF scales roughly as the fifth power of the relative effective index $n_{eff}^5$ [49]. We calculated the PF of the F-MSM waveguide following the approach and approximations of Ref. [49]. In this case the PF is defined by waveguide properties, in particular, field profile and dispersive curve. Results show that PFs above one hundred can be achieved with sufficiently thin metal layers (Fig. 2b).

While the relative effective index of the Sa mode increases with decreasing the metal thickness, the confinement factor in a F-MSM waveguide has the opposite tendency. Both confinement factors defined via $E^2$ and $P_x$ either in the waveguide core only or in the waveguide core and two metal plates are decreasing (Fig. 3a). Such behavior is abnormal for a conventional optical waveguide, where the field confined in a higher index structure causes higher relative effective index. To solve the contradiction we calculated the confinement of $P_x$ and $E^2$ only in two metal plates (Fig. 3b). $\Gamma(P_x)$ in two metal plates is also decreasing. However, $\Gamma(E^2)$ is increased and it agrees with the increase of the relative effective index. Nevertheless, the magnitude of these changes is much lower than magnitude of changes in waveguide's core and surrounding SiO₂.

Fig. 4 shows a comparison of $P_x$ and $E^2$ profiles for two thicknesses of metal layers. As the thickness decreases, for each metal stripe, the field is localized closer to the metal surfaces. This, in turn, generates less coupling of the two metal stripes and, therefore, less field is confined inside the waveguide core. These changes in the field distribution inside the waveguide core are more pronounced than at the interface and as a result they have more influence on the confinement factor.

## 4. Absorption coefficient and modulator's performance

Although the increase of the effective index is a desired effect, doing it by thinning the core and metal claddings has a drawback. The fields penetrate more into metal and cause an increase of losses in the system. As a consequence, the absorption coefficient of the F-MSM waveguide with thinner metal layers is higher (Fig. 5a). Therefore, loss compensation requires higher gain. We calculated the critical gain needed for the full loss compensation in F-MSM waveguides with thin metal layers. In Fig. 5b we plot the relative effective index values that can be achieved with certain gain levels and core thicknesses. As it can be seen, thinner cores allow higher effective indices. The required material gain is feasible for InP-based semiconductors utilizing either bulk material properties or a stack of quantum wells [35,50].

We considered that the active core possesses gain in the on-state and absorption in off-state with the same absolute value of ε". The logarithmic extinction ratio (ER) *ER* =

$8.68 \cdot (\alpha - \alpha_{on})$ shows how deep we can modulate a propagating wave per waveguide's unit length. We performed calculations of the extinction ratio of the F-MSM waveguide for $d_1 = 30$ nm and $d_2 = 50$ nm, and various thicknesses of the metal layers. Calculations for the uniform gain core (Fig. 1a) show that for the same amount of gain, the thinner active core exhibits better performance (Fig. 6). Thus, ER = 5 dB/µm and higher can be achieved. The results for symmetric (Fig. 6a) and asymmetric (Fig. 6b) structures are similar. It shows that at least one thin metal film is needed to achieve high performance.

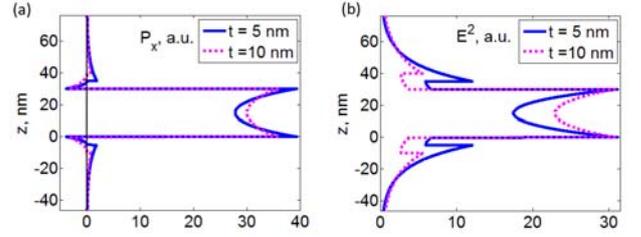

Fig. 4. (a) Poynting vector $P_x$, along the propagation direction, and (b) $E^2$ field profile in the symmetric waveguides with $d = 30$ nm and different metal thicknesses: $t = 5$ nm and 10 nm. All profiles are normalized such that the total integrals over whole cross-section are the equal.

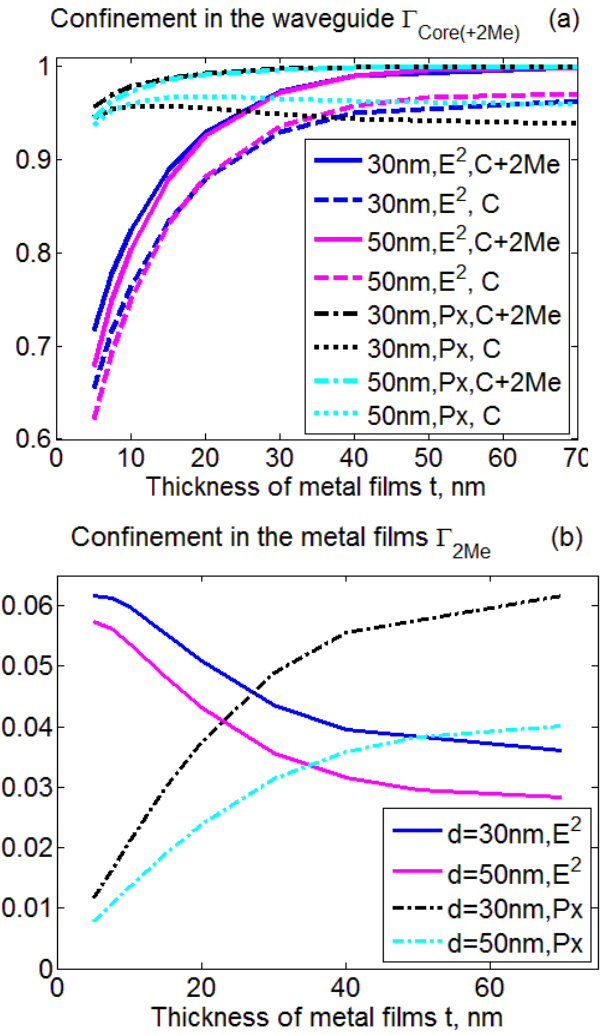

Fig. 3. (a) Confinement of electric field ($E^2$) and Poynting vector ($P_x$) for symmetric waveguides with core $d = 30$ and 50 nm inside the core (C) or inside core and two metal plates (C+2Me). (b) Confinement for the same waveguides but inside only two metal plates (2Me).

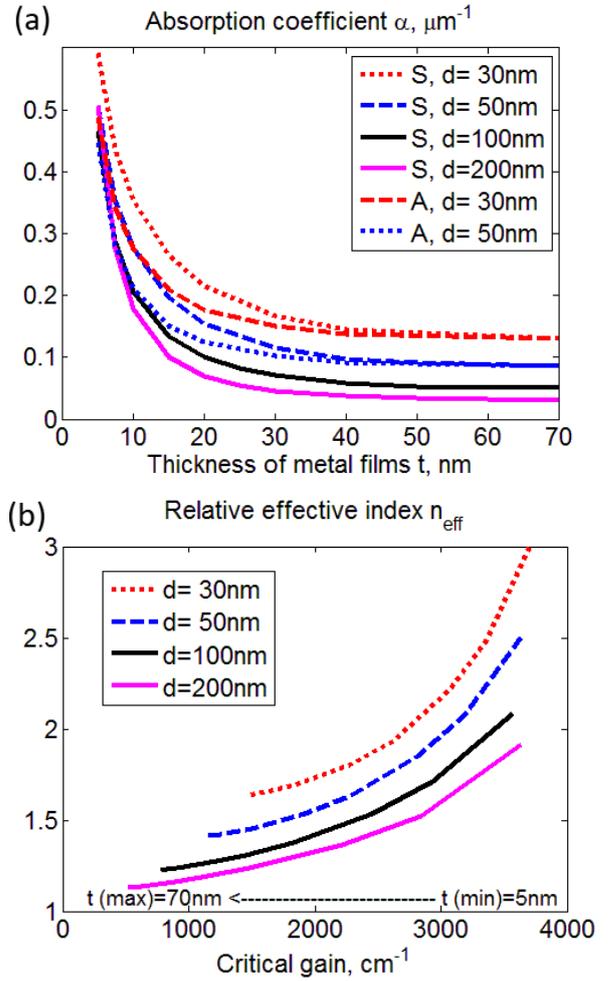

Fig. 5. (a) Absorption coefficients of F-MSM waveguides (symmetric (S) and asymmetric (A) structures). It significantly increases for thin metal layers along with the increase of the relative refractive index (see Fig. 2a). (b) Relative effective index that can be achieved in a symmetric waveguide with the provided level of the critical gain (metal losses are fully compensated). The metal layer thickness is varied along each line.

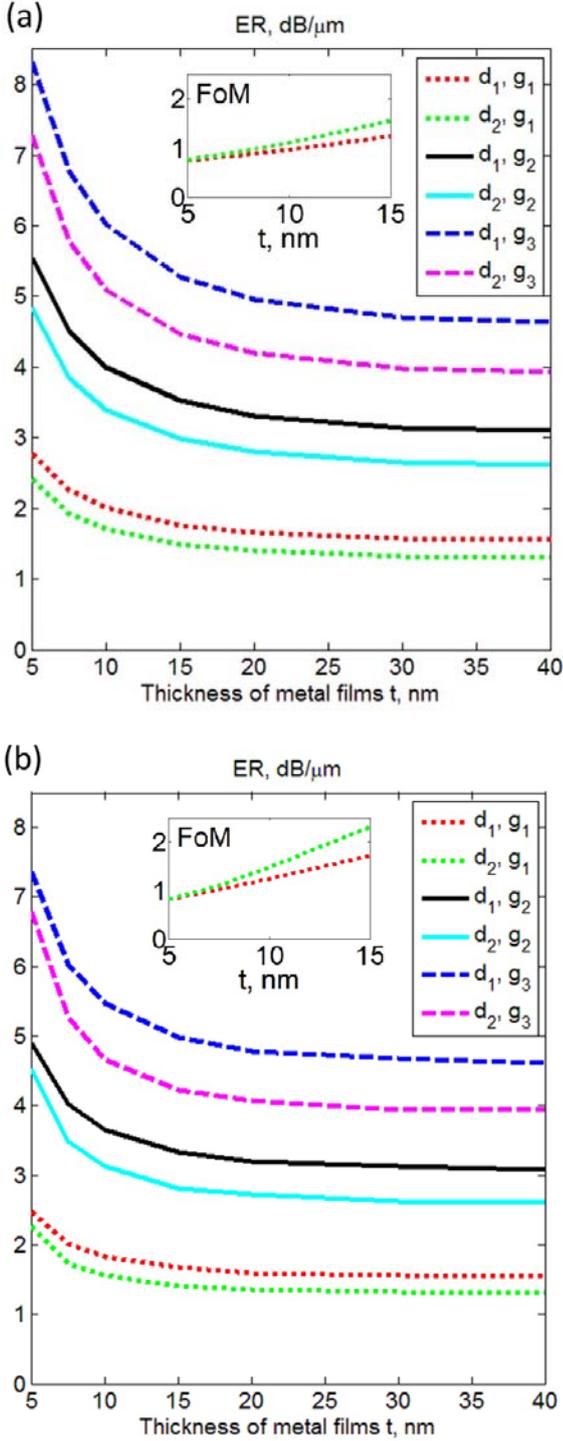

Fig. 6. Extinction ratio of F-MSM waveguide ($d_1 = 30$ nm and $d_2 = 50$ nm) for different gain values of the InP-based active core: $g_1 = 1000$ cm$^{-1}$, $g_2 = 2000$ cm$^{-1}$, and $g_3 = 3000$ cm$^{-1}$. (a) Thickness of both metal films is varied simultaneously. (b) Thickness of one metal film is fixed ($t_1 = 70$ nm) and another is varied. Insets: FoMs of two cases respectively.

To characterize the performance of the modulator, we introduce through the absorption coefficients the figure of merit (FoM): $\text{FoM} = (\alpha - \alpha_{on})/\alpha_{on}$. In contrast to ER, FoM increases with increasing of thickness of metal films (Fig. 6, insets). It is due to the significant decrease of $\alpha_{on}$ for thicker metal layers. The FoM of this MSM device is several times higher than for MIM plasmonic modulators with the silicon nitride and indium tin oxide multilayered core [39].

## 5. Influence of n- and p-doped layers

To analyze a realistic structure in case of electrical pump, we consider a F-MSM waveguide with n- and p-doped layers (Fig. 7a). These layers are required to inject carriers into the gain core. In calculations we consider 10-nm-thick layers with permittivity $\varepsilon = 12.46$. In our approximation, n- and p-doped layers do not possess any gain or loss. We study the same parameters of the waveguide as for the uniform waveguide core.

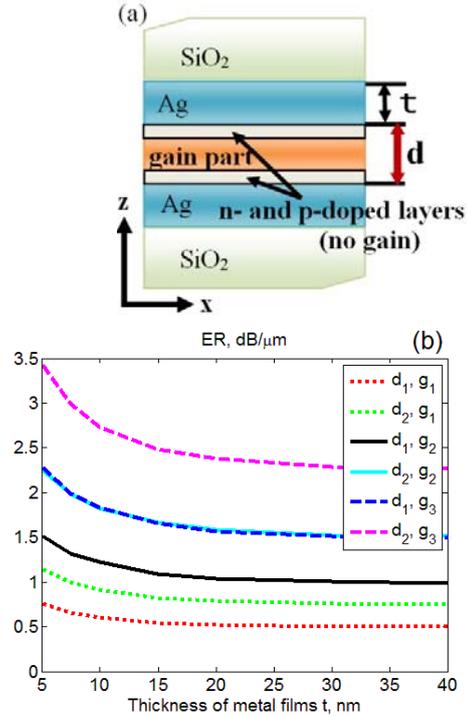

Fig. 7. (a) F-MSM waveguide with n- and p-doped layers and metal films of equal thicknesses $t$. Core thickness $d$ includes the gain part and both doped layers. (b) Extinction ratio of the waveguide ($d_1 = 30$ nm and $d_2 = 50$ nm) for different gain of the InP-based active core: $g_1 = 1000$ cm$^{-1}$, $g_2 = 2000$ cm$^{-1}$, and $g_3 = 3000$ cm$^{-1}$. Calculations are performed for 10-nm-thick doped layers.

However, in case of the F-MSM waveguide with the n- and p-doped layers, the sample with total core thickness $d_1 = 30$ nm gives lower performance than waveguide with $d_2 = 50$ nm (Fig. 7b). In this case, the extinction ratio is lower, ER = 1-2 dB/μm. According to the calculated field distribution (see Fig. 4b), in the core of the waveguide the field is strongest next to the metal interface, and so active properties of the materials are the most pronounced there. In the same time, the doped layers are placed in that part and thus limit the effect of the gain material.

Further, we analyze a waveguide with a 5-nm-thick metal layer and varied core thickness $d$. Absorption coefficients show a non-monotonic dependence with a minimum around $d \approx 100$ nm (Fig. 8a). However, the ER of the system is monotonically rising with increasing of the waveguides core, i.e. the thicker the core the larger is ER (Fig. 8b). In other words, taking into account the n- and p-doped layers thicker waveguides are preferable. These results are consistent with ones obtained for the MSM waveguide with infinitely thick metal layers [35].

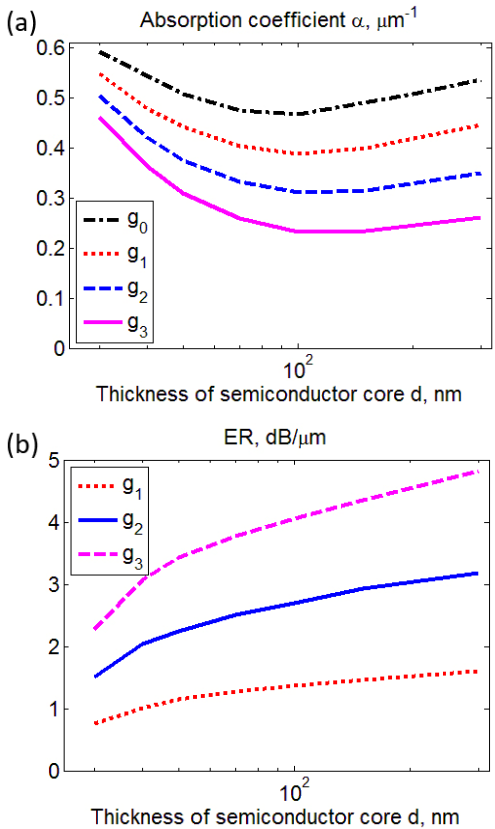

Fig. 8. (a) Absorption coefficient and (b) extinction ratio of F-MSM symmetric waveguide with n- and p-doped layers at various gain values: $g_0 = 0$, $g_1 = 1000$ cm$^{-1}$, $g_2 = 2000$ cm$^{-1}$, and $g_3 = 3000$ cm$^{-1}$. Thickness of metal layers is fixed, $t = 5$ nm. Calculations are performed for 10-nm-thick doped layers.

## 6. Finite-length F-MSM waveguide

To study wave propagation in such F-MSM waveguide we performed frequency-domain simulations using the commercial software CST Microwave Studio [51]. Effectively 2D system layout is presented in Fig. 9a. It consists of an active and two identical non-active parts. A mode is excited at the left port and sequentially propagates through the first non-active part, active part and the second non-active part. As far as the wave propagates it is attenuated and reflected at the interfaces between the active and non-active parts. We calculated the dependence of the reflection and attenuation values with the length of the active part $L_a$ (Fig. 9b,c). We performed calculations for two systems. The first one is a waveguide where both active and the non-active parts of the core have Re($\varepsilon$) = 12.46 (referred as "U"-waveguide in Fig. 9b,c), thus providing low reflectivity at the interface (see Fig. 9b,c). In the second, the waveguide core material of the non-active part differs from the one of the active part (referred as "C"-waveguide). In this case we considered the permittivity of the non-active part as $\varepsilon_{na} = 4$.

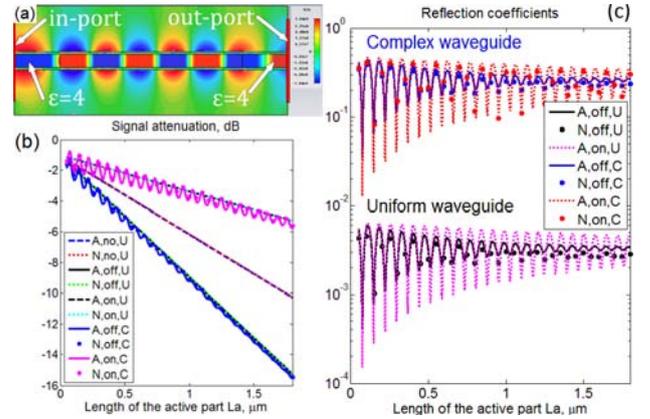

Fig. 9. (a) Instantaneous electric field distribution in a complex F-MSM waveguide with active and two non-active parts. Thicknesses of metal layers is $t = 5$ nm and core $d = 30$ nm. Logarithmic attenuations (b) and reflection coefficients (c) in different structures. Notations: "A" stands for results obtained as analytic solutions based on the transfer-matrix approach; "N" stands for numerical simulations; "C" for complex waveguides in contrast to uniform ones "U"; "off" for the core with absorption $\varepsilon'' = 0.174$; "on" for the active core with $\varepsilon'' = -0.174$ (corresponds to gain $g_2 = 2000$ cm$^{-1}$); "no" for the passive core $\varepsilon'' = 0$ (neither gain, no loss).

We also employ the transfer-matrix approach [52] to find an analytic solution for the mode propagation reflection and attenuation values. Complex propagation constants in the active ($\beta_a + i\alpha_a$) and non-active ($\beta_{na} + i\alpha_{na}$) parts of the waveguide are derived from dispersion relations and further substituted in expressions for propagation in layers with different thicknesses. The effective mode indices are defined as $n_a = (\beta_a + i\alpha_a)/k_0$ and $n_{na} = (\beta_{na} + i\alpha_{na})/k_0$, where $k_0 = 2\pi\nu/c$ is the wave number in vacuum. The whole transmission matrix is a product of transmission matrices for single interfaces and propagation matrices of single layers. We consider the single interface transmission and reflection coefficients in the transmission matrix defined using the classical formulae where the refractive indices are replaced with the effective mode ones: $t = 2\sqrt{n_a n_{na}}/(n_a + n_{na})$ and $r = (n_a - n_{na})/(n_a + n_{na})$.

Numerical and analytical results coincide well (see Fig. 9b,c). This agreement leads to the conclusion that the analytic transfer-matrix approach using effective mode indices can be applied for plasmonic waveguide systems. In the same time, it shows that no other waves are excited at the waveguide interface so the symmetry of the Sa mode is preserved.

## 7. Finite-width F-MSM waveguide: 3D case

To compare 2D approximations with a more realistic structure we analyzed a three-dimensional (3D) F-MSM waveguide of finite width $w$ in the y-direction (see Fig.1a). We perform numerical simulations for various widths from 20 nm to 3 μm studying two cases: the first layout with the sharp corners and the second one with the corners rounded with 2.5-nm-radius curvature. We considered on-, off- and insulator-state (Fig. 10). Insulator-state corresponds to core without gain ($g = 0$). For relatively small widths ($w \leq 500$ nm) both propagation constant and attenuation coefficient differ significantly from results obtained in the 2D approximation (Fig. 10). The trend shows that the propagation constant can be increased but with the price of an increase in losses. Confinement of the mode in two dimensions causes even higher field localization and thus increasing losses (Fig. 11a).

With the increase of the width, the parameters tend to those of the 2D approximation with the infinite width. However, it can be noticed from Fig. 10 that the calculated values do not converge exactly to 2D ones. Similar results can be found in [35,38] for other configurations of MSM waveguides. Moreover, the effect was studied experimentally for a metal stripe waveguide and was ascribed to contribution of radiation damping due to field scattering on edges [53]. Indeed, at the waveguide edges the fields are significantly different, even for wide waveguides, from the ones in the middle of the waveguide that can be considered similar to the ones of a 2D waveguide (see Fig. 11b for a waveguide width $w = 3$ μm). Nevertheless, the quantitative difference between the 2D and 3D results with $w \geq 500$ nm is not large: 0.7% for propagation constant and 1.5%, 3% and 7% for attenuation coefficients in off-, insulator-, and on-states, respectively. We conclude that 2D analysis can be applied for F-MSM waveguides with widths bigger than 500 nm.

Difference in sharp and rounded corners is noticeable for small $w$, but hardly visible with large $w$. As the result, for large $w$, the rounding of the corners does not influence the waveguide's parameters so does not increase the convergence to 2D values.

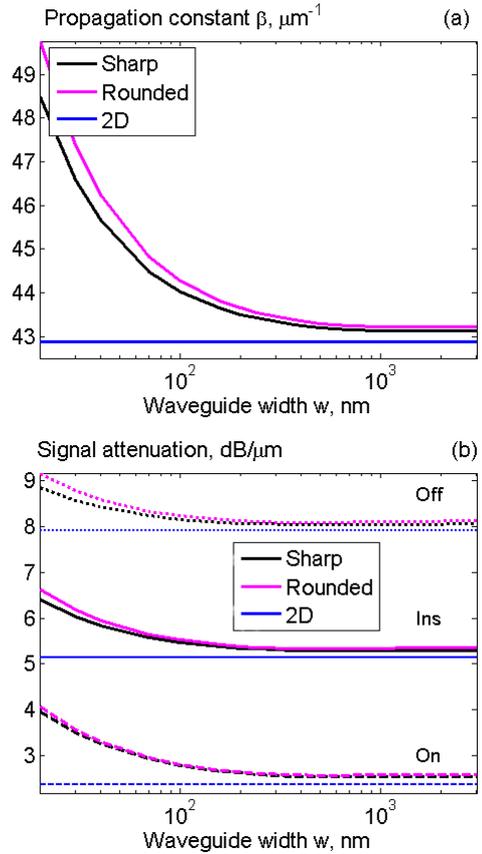

Fig. 10. Characteristics of the F-MSM waveguide of finite width $w$. The thickness of metal layers is $t = 5$ nm and core $d = 30$ nm. (a) Propagation constant and (b) signal attenuation in the F-MSM waveguide for various waveguide widths.

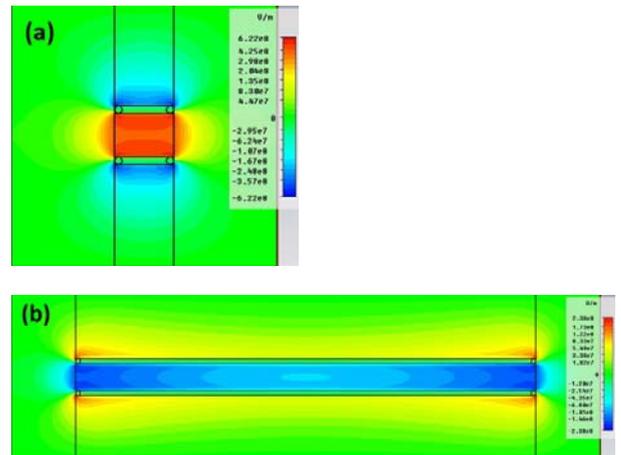

Fig. 11. Mapping of instantaneous electric field distribution in the cross-section of the F-MSM waveguide with $w = 20$ nm (a) and 3 μm (b).

## 8. Discussion and conclusions

Due to the strong field localization MSM structures are featured as ultra-compact devices. For instance, an ultimate size semiconductor nanolaser has been recently designed based on a single quantum well sandwiched between thin silver layers [27]. Along with [35] it shows advances to utilize ultra-thin gain nanostructures embedded in metal layers.

The fabrication of InP-based semiconductor devices brings several challenges. In particular, crystalline structure can be grown only on another InP-based material. Thus additional techniques, e.g. flip-chip and bonding, to place the active layer between the metal layers are needed. However, namely the InP-based devices provide the highest gain and consequently have the highest potential in compensating metal losses. It should be also mentioned that the considered structure is completely planar and, as the result, it does not contain any features inside waveguide, which can significantly complicate the fabrication process.

Including additional layers or designing a periodic structure could enhance the performance of the system even more (see for example [54]). In this case, instead of InP based semiconductors alternative gain materials, like colloidal quantum dots or dye solutions, may be used. In contrast to crystalline semiconductors, colloidal quantum dots and dye solutions can be deposited layer by layer and, thus, arranged in multilayered structures.

In some calculations we considered very thin metal layers, in particular down to 5 nm. At such small thickness the metal properties can be different from the bulk ones. Moreover, non-local and quantum effects start to play a role. Nevertheless, we believe that our approximation shows the general qualitative tendency even if the actual numbers may be quantitatively different.

The structure can be controlled both electrically and optically. The electrical pumping can be realized using the metallic plates as electrodes, thus one can say that it is intrinsically built-in. However, in this case we need the n- and p- layers and as the result the device's performance is decreased. Furthermore, an additional speed limitation comes from the RC-circuit set up by the structure. In contrast, there is no need of doped layers and no frame of RC-circuits in the layout with the optical pumping, as studied materials form Schottky junction on the interface metal-semiconductor. Thus, more efficient and faster modulation can be realized. However, the optical pumping scheme requires a more sophisticated layout to get access to a source and deliver the modulating signal.

We have showed that utilizing thin finite metal films in a MSM waveguide can increase the relative effective index in several times compared to a MSM waveguide with thick layers [35]. The relative effective index increase results also in increase of the Purcell factor and thus enhancing the spontaneous emission rate. The main control mechanism is to modulate stimulated emission of semiconductor. Therefore, from one side, the higher speed of spontaneous emission the faster stimulated emission can be switched off. Thus, achieved Purcell factor (above 100) may lead to orders of magnitude increase of the modulation speed.

On the other side, a higher Purcell factor leads to increased spontaneous emission thus extra noise in the device. Nevertheless, for device's analysis we can neglect the increase in the noise level and consider that the signal is significantly higher. This is due to two factors. First, in usual photonic devices, noise is accumulated along the waveguide (tens or even hundreds of microns) and results in undesirable effect. However, the active part of the considered plasmonic modulator is only several microns long and several tens of nanometer thick, and does not allow rising of the noise level. Second, noise is subject to plasmonic losses in the waveguide and thus it is greatly diminished.

The high Purcell factor significantly increases the current density that is required for achieving loss compensation in the waveguide. It was pointed out as the main obstacle for utilizing the MSM structures [50]. Indeed, very high current intensively heats the structure. However, again, because of small dimensions of the structure and its crystalline nature, efficient outward heat transfer can be implemented, for example, with the help of thermoconducting materials as the waveguide claddings.

The suggested plasmonic modulator can be part of a metal-insulator-metal waveguide or coupled to another type of plasmonic waveguide with high mode localization. There are various possibilities to couple metal-insulator-metal waveguides to conventional photonic waveguides [54-57]. Similar concepts can be utilized for our F-MSM structure. A detailed study of various couples was not the aim of the present work.

In summary, we proposed to exploit an MSM waveguide with finite-thickness metal layers for modulation of plasmonic waves. In this work we were interested only in the symmetric low energy dispersive branch as it gives the highest field confinement, propagation constant and absorption coefficient. The absorption coefficient is even higher than the ones of the MSM waveguide with thick metal layers. However, particularly this property provides higher extinction ratio of the modulator based on the F-MSM waveguide. The thinner the metal layers the higher the extinction ratio, namely 3-8 dB/µm with a feasible gain level. We believe that, using this property, the plasmonic F-MSM structures can outperform earlier proposed layouts.

The main advantage with respect to the silicon based modulators [58,59] is the footprint of our device. In particular, it can show 10-dB modulation depth with a footprint of about 2x0.5 µm$^2$ (length x width). We have not yet investigated the bandwidth characteristics but the modulation concept is non-resonant, so the operating bandwidth can be high. Such investigation is part of future work. Modulation with speeds up to 500 GHz is feasible, as tight mode confinement between metal plates increase the rate of active material response [50,60].

**Acknowledgments**

Authors are very grateful to Kresten Yvind and Andrei Andryieuski for fruitful discussions. R.M. and A.V.L. acknowledge partial financial support from the Danish Research Council for Technology and Production Sciences via the THz COW project.